# Conversational (dialogue) model of quantum transitions

Pavel V. Kurakin (1), George G. Malinetskii (1), Howard Bloom (2)
(1) Keldysh Institute of Applied Mathematics, (2) New York University

We introduce an original model of quantum phenomena, a model that provides a picture of a "deep structure", an "underlying pattern" of quantum dynamics. We propose that the source of a particle and all of that particle's possible detectors "talk" before the particle is finally observed by just one detector. These talks do not take place in physical time. They occur in what we call "*hidden* time".

Talks are spatially organized in such a way that the model reproduces standard quantum probability amplitudes. This is most obviously seen if one uses R. Feynman's formulation of quantum theory. We prefer the "physical level" of mathematical strictness in describing our model.

The model was initially designed without any background except R. Feynman's many – paths approach. But it later became apparent that the suggested mechanism is highly generalizable, and may apply to a wide spectrum of self – organized systems, including living systems. Stories about such systems are attached in appendices.

Comments: 16 pages, 10 figures

## 1. Introduction

As a 1$^{st}$ step, we confine our treatment to radiation of light quanta by unmoving charges only, and show how that the "underlying picture" we introduce for quantum dynamics leads to exact quantum electro - dynamical amplitudes. The description is at a "physical" level of mathematical strictness.

Our basic idea is based on Feynman's many – paths formulation of quantum mechanics [1, 2]. We suggest that path integrals are not simply a mathematical technique. Instead these integrals *can* reflect "talks" that take place between the source and many detectors. Talks imply many passes of signals in both directions along each path. It resembles very much transactional interpretation of quantum mechanics by Prof. John Cramer [3].

These talks *physically* accomplish spatially distributed computation of probability amplitudes for different possible detectors. All this computation has no deal with physical time and is processed in hidden time. We provide a clear and physically strict description of relation between hidden time and physical time and show that our reconsideration of the notion of physical time is fully relativistic.

Next, we show that though we start from unmoving charges, our model *intrinsically* contains an exchange of momentum (between the source and the detector) and spatial displacement of interacting charges.

At last, we provide our hypotheses about how our model *can* possibly provide gravity, quarks and antiparticles for free.

## 2. Basic analogy: "bees in the hive"

Consider how a hive of bees makes up its collective "mind". The majority of the bees who work outside the hive are foragers. They fly in packs, shun independence, and bring home the goods – the pollen, nectar, or water – that the hive needs most at the time. The conformist, sheep-like nature of the foragers maximizes their focus on harvesting. It maximizes their efficiency in mining the resource of the moment. But it minimizes their odds of feeling out new possibilities for the hive.

Meanwhile a tiny number of bees have a very different job. They are explorers, restless, solitary wanderers who fly in seemingly random patterns, checking out the landscape for new sources of excitement, for new flower patches or for new sources of water. Worker bees want to harvest the goods the hive needs most. But first they need the guidance of a good group decision about the most useful place to fly (Fig. 1a). How does that decision get made?

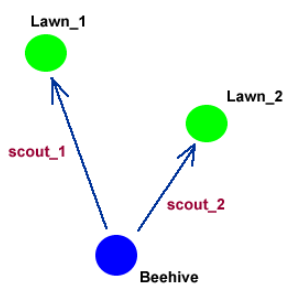 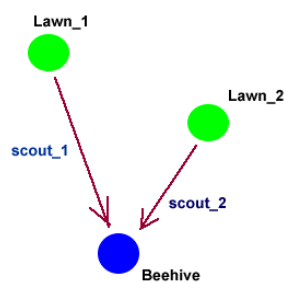 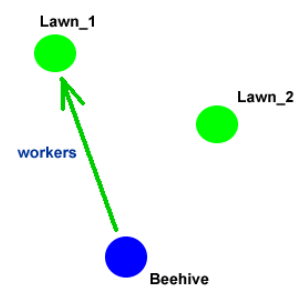

*Fig. 1a*        *Fig. 1b*        *Fig. 1c*

A scout who is fortunate discovers a flower patch or water puddle she feels is promising. Then she comes back home and starts to advertise her find (Fig. 1b). As Karl von Frisch explained in his classic *The Dance Language and Orientation of Bees* [7], scout bees promote their discoveries by dancing special figure-eight-shaped dances. Those figure-eights explain the direction of the find and the difficulties or the extra boost given by headwinds and by tailwinds along the way. Most important, the length of a scout bee's dance gets across her "enthusiasm," her sense of the value of the treasure she has discovered.

Worker bees attentively "listen" to these advertisers. They gather in crowds around four or five competing dancers, absorbing the message of their figure eights and of the performers' enthusiasm, her endurance.

If a dancing scout just won't give up, a few of the foragers in her audience will catch a bit of the dancer's enthusiasm and go out and check her report. If they're impressed by what they find, they come back and join the dance. Eventually one bee manages to gather the biggest audience and the largest number of background dancers. That's when the hive makes up its mind. The foragers go off in a pack to mine the flower patch the winner of the dance contest has marketed and promoted so successfully [8]. (Fig. 1c).

Meanwhile, this is a two-way conversation between senders and receivers, between a source and a bevy of detectors. The scout bees are not the only competitors whose input counts. A competition has also occurred between the receivers of the bees' attentions – the flower patches that offer their nectar and pollen. Flowering plants – angiosperms – use their nectar and pollen to motivate bees to spread the plant's pollen to other sexually-receptive flowers nearby. The petals of a flower and the flower's heart are temporary billboards erected to rouse a bee's interest. The flower patch that advertises most successfully stands the best chance of catching a bee-hive's attention and of becoming the next popular flower patch, the next foraging target, of the day.

The analogy between quantum particles and bees is suggested by Howard Bloom. It is a qualitative analogy only. It shows only how one of many possible detectors can in principle be chosen via talks between the source and the detectors.

See more on the bees' talks in *Appendix A*. The next chapter gives more exact explanation of the model.

## 3.    Talks in hidden time

We remind the reader that we assume unmoving charges exchanging a single photon. We suggest that the space[1] consists of discrete nodes, some of which are "empty" or "void", and some contain charged particles, emitting (and absorbing) photons. Radiation and absorption of a photon is accomplished through the following steps.

1. The source charge S sends scout signals to the nearest void nodes. Each node which receives a scout signal sends the same signal to its nearest neighboring nodes. In such a way scout signals reach *all* existing possible detectors $D_1$, $D_2$, $D_3$, $D_4$,... and so on (Fig. 2a).

Perhaps it could be more appropriate and adequate to formulate the model in terms of *ribs* that connect nodes, or in terms of both nodes and ribs.

A rib can be *void*; or it can be signed by some signal. We have 3 (or may be 4) kinds of signals in our model: scout signals, query signals and confirmation (or refusal) signals. So, the 1st step marks all the ribs in *all* the Universe, ribs that were previously void, with scout signals, pointing from the source **S** to all possible detectors.

- For the next iteration, imagine that each rib marked by scout signal is also characterized by a *phase variable* φ. While transferring from a rib to the next rib, the phase variable "rotates": $φ_{n+1} = φ_n + Δφ$; $Δφ = \{2π(l/λ)\} \bmod 2π$. Here l is the space distance between centers of ribs, and λ is the wavelength of the photon considered.

This implies that a void node should transfer scout signals by sequence. In other words, scout signals must not interact in void nodes. Or, they move linearly through the space, with no influence of other scout signals.

2. When scout signals have reached a charged node $D_i$ *by all possible paths*, phases variables of scout signals are summed <u>as vectors</u> of unitary length and direction determined by φ: **Φ**$_D$ = Σ**φ**$_{in}$. It is standard quantum

---

[1] Perhaps, more general description should include *space – time*. But from the following text one will see that using the space only is enough to build relativistic – covariant model for photons.

electro - dynamical amplitude for the photon to be absorbed by detector $D_i$.

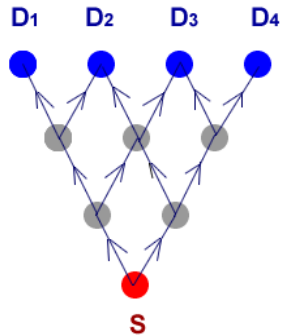
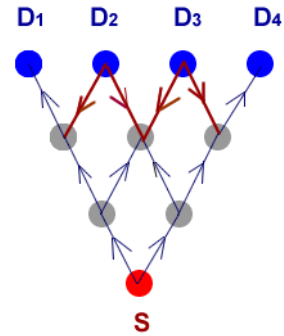

Fig. 2a          Fig. 2b

The charged node stores the total sum only and increases its value consequently with each incoming scout signal. A very interesting question here arises: how the charged node "knows" that all possible signals have already come in, and it's time to stop counting the sum?

Intrinsic need to *stop* phase summing seems to complicate the model. Actually it occurs that this need is also tightly connected to momentum exchange and spatial displacement of interacting charges (see section 5).

3. After the charged node has counted its amplitude $\Phi_D$, it generates *query signal* and broadcasts it worldwide (Fig 2b). Query signals follow the trace of scout signals. Broadcasting of query signals resembles how the source node broadcasts scout signals. I.e., the signal is transferred by void nodes from a rib to rib (both previously marked by scout signal).

4. The phase variable is not introduced for query signals. Instead the invariable intensity $I = |\Phi_D|^2$ is considered. At each void node a *lottery* happens. Incoming query signals compete to be transferred next (Fig 2c). Each query signal has a chance to win lottery, while relative probabilities *relate as intensities*.

This must provide that *each* possible detector $D_i$ has relative probability $I = |\Phi_D|^2$ to be a winner at *each* void node.

When the lottery has finished, the winner signal propagates next, while all ribs containing loser signals change to void state (Fig. 2c).

Finally query signals reach the source. Source node **S** makes a lottery just like void nodes and the only query signal survives. Since the rule "a single incoming query survives at each node" works through whole Universe, it's clear that there emerges a polyline path connecting the source **S** and some single detector $D_i$ (fig. 2d). A *confirmation signal* runs from **S** to that $D_i$. When it arrives, this means the photon *is detected* at $D_i$.

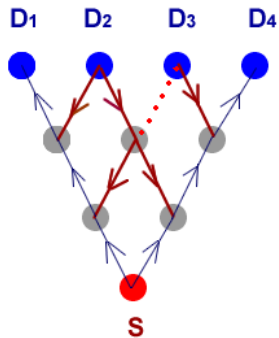
Fig. 2c

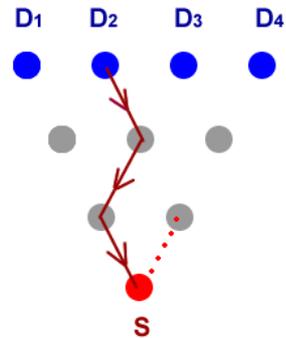
Fig. 2d

5. Let a void node has only void ribs but a single one with incoming query signal. This means that all copies of this signal, previously copied "down", are losers now. This rib is marked to be void, too.

This means propagation of "refusal" signal in direction from **S** to loser detectors. For analogies in living cell dynamics see *Appendix B*.

## 4. Sewing of hidden time and physical time

(a) All signals introduced above propagate in hidden time. One can treat "hidden" as simply "not physical". The physical time does not tick when such signals move. Actually hidden time is not a time at all. Instead it is a mathematical abstract, which supports the *dynamics* of nodes and ribs.

We suggest accepting, that physical time has a local meaning; it ticks at a definite detector node when a confirmation signal arrives to it.

To make this definition consistent with experiment, we suggest accepting that actual detecting of physical time we make in *any* experiment can fundamentally be expressed in terms of counting some elementary events.

Subsection (b) below explains what exactly we suggest to assume as *locally detected* physical time; subsection (c) shows that our model is consistent with this way of detecting time; subsection (d) argues that the procedure we define in subsection (b) (and thus our model as well) is *relativistic covariant*.

(b)     Imagine we emit a photon from source **S** and detect it at detector **D** (Fig. 3). Let we start counting time by clock **C** placed near **D** simultaneously with photon's departure from **S** (in resting referential frames). In other words, emitting of a photon is synchronized with the switch of the clock **C**. When the detector **D** works, it switches the clock **C** off.

Clock **C** can be any highly stable oscillatory or stationary process. Actually, the clock can combine both. Say, a photodiode **P** can supply stable electric current generated under radiation of highly stable laser **L**.

In reality we detect total energy absorbed by laser **L**. Still, if we know the line length of the laser, effectively we detect the number of light quanta emitted by laser.

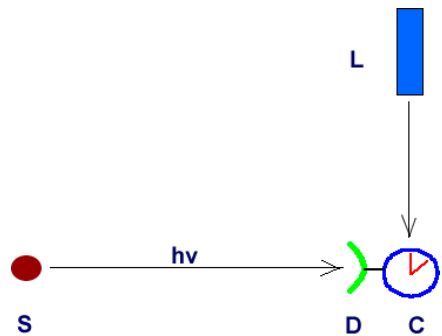

*Fig. 3*

Here we suggest that fundamentally, with maximal idealization, roles of detector **D** and clock **C** can be combined in a single atom **A** provided the photon emitted at **S** has the same frequency (energy) as laser **L** emits (we imagine idealized, infinitely narrow line), and atom **A** has the same transition frequency.

Imagine we relax the atom **A** somehow after it absorbs a photon from **L**; either we can manage to substitute a new unexcited atom to the place of **A**. Next, it's important that each absorbed photon transfers a vertical momentum to **A**. On the other side, the photon from **S** transfers horizontal momentum.

This means we are fundamentally able to distinguish photon coming from **S** from photons coming from **L**. In such case we can *define* the time elapsed at **A** (**A** = **D** = **C** here) during photon's flight (from **S**) as the *number* of absorbed photons from the laser **L**.

It's interesting that the role of momentum recoil in time detection was previously mentioned by H. Salacker and E. P. Wiegner [4]. Authors pointed that momentum recoil of the clock is an inevitable feature of any time detecting procedure.

We suggest here that momentum recoil is not just a *feature* of clocks, but is a *necessary* condition of any time detecting procedure.

Of course, a single atom clock is not precise anyway, because the detection of photons from both **S** and **L** is not deterministic: it obeys quantum probability amplitudes. Still we believe that a single-atom clock is able to reflect the fundamental processes underlying any mechanism for detecting physical time.

(c) The "talks protocol" introduced in section 3 is fully consistent with the time detecting procedure introduced above. Imagine for simplicity that we use a single path. In other terms, there's no summation: a single scout arrives from each source. In this case the scout signals coming to any detector stand in a queue (Fig. 4). We need to add one more single rule:

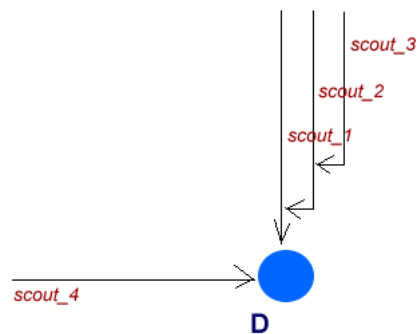

*Fig. 4*

Detector starts developing any scout signal, i.e. sending appropriate query signal and "waiting" (in hidden time) for the reply *only* after it has finished developing previous (in the queue) scout signals.

In this case the whole model stays consistent with experiment, i.e. with the constant light speed *c*. While the scout signal *scout_4* travels from source **S** detector **D**, scout signals from laser **L** come to **D** and consequently join the queue.

If *scout_1* comes simultaneously (in hidden time) with the departure of *scout_4* from **S**, then at the instant when *scout_4* arrives to **D** the size of queue is proportional to distance between **S** and **D**.

We suggested here that speed of all scout signals is the same in hidden time. We also suggested that *simultaneity* in hidden time (space – time, more generally) is absolute, just like in Galilean world.

So, the time of photon flight from **S**, when measured as the number of photons from laser **L**, is proportional to distance between **S** and **D** with constant factor.

(d)  Proving Lorenz covariance of defining physical time through counting of elementary events like absorption of a photon is rather trivial (at physical level of strictness).

We suppose that *all* light quanta from the laser **L** are absorbed by atom **A**. Of course, it *is not* literally true for single atom but it is true for classical detector when normalized to number of atoms in detector. So we need to count light quanta emitted from **L**.

Exact quantum – electro dynamical formula for light intensity (which equals to number of emitted light quanta per unit of time) [5] gives that stationary laser intensity is proportional to $Q$ of resonator cavity, which in turn is reciprocal to length of cavity. If we intend to assume different referential frames, it would be more correct to use light path within cavity rather than length of cavity.

That is, time to emit one photon (we can detect it as number of some other, more *frequent* elementary events) is proportional to path length within cavity. Since the number of full paths does not change from one referential frame to other (it is a scalar), we need to examine how the single full path or the oscillating period changes.

Let the clock (system **L** + **A**) moves with velocity $v$. From figure 5 it's clear that

$$c^2 t^2 = v^2 t^2 + c^2 \tau^2 \quad (1).$$

In (1) $t$ is time needed by light to make one pass, when observed in resting referential frame, and $\tau$ is the time for one pass observed by clock itself. In other words, $\tau$ is own time of the clock.

From (1) we directly get that $t = \tau\sqrt{1 - \dfrac{v^2}{c^2}}$.

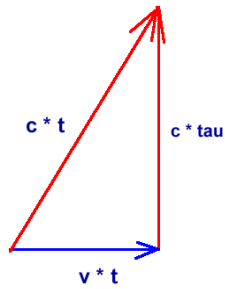

*Fig. 5*

In subsection (d) we have argued that the way to detect physical time we suggest is consistent with relativity. In subsection (c) we showed that the protocol of conversational model we suggest is consistent with such a way to detect physical time. So we are brave to conclude that the whole our model is consistent with relativity, though we still talked in terms of unmoved charges only.

## 5. More physical features emerge for free

1. Entanglement

Entanglement is, perhaps, the most obvious phenomenon explained by the hidden time model. Query signals that return to the source of entangled particles accomplish correlation at the very source. Still this co-ordination happens in hidden time, not physical time, which removes possible conflicting with Bell's theorem.

This implies that detected (at future) values of correlated quantity shared by entangled particles *do not* "exist" just at the emitting instance. They do not exist in physical time; still they are determined in hidden time.

2. Momentum transfer

As we pointed before, we start from resting charges. But the time detecting technique we introduce requires that momentum recoils from the horizontally and vertically moving photons we are trying to distinguish. Thus our model *intrinsically needs* momentum transfer, and it *can possibly* provide it for free as follows.

As we noted above in section 3, our model needs to provide a mechanism for detecting charge **D** to stop counting the phase variable φ of incoming scout signals.

We suggest, that this counting and its halt can be constructed, from general principles, in the following way. Scout signals coming from one source propagate by close paths, and thus the values of the phase variable as the scout signals arrive at a detecting charge differ slightly: $\Delta\varphi \ll 1$.

On the other side, a scout signal coming from some other source will, most probably, essentially differ from the scout signal which came before it: $\Delta\varphi \sim 1$.

When such a new scout arrives, the node determines that it's from a new source and stops counting the phase from the first source. The summary $\Phi_D$ is put in the queue as we described in section 4(c).

The queue has to be spatially organized in this or that way, and we propose that this organization takes place in hidden space, indeed. We suggest that this *can effectively lead* to the spatial displacement of the detecting node from its starting position.

We will risk suggesting some even more profound explanations that emerge for free from the conversational model.

3. "Antiparticles"

Note that void nodes and physical nodes (charges) have a common feature: they provide a locus for lotteries. They also differ in the manner in which they pass through or reflect hidden time signals. We suggest that void nodes may effectively "consist" of two physical nodes with opposite charges. This pairing of opposite charges is analogous to electron – positron pairs in the vacuum.

4. "Quarks"

A physical node, i.e. a charge, generates 3 kinds of signals in our model. There can be 3 kinds of different "1\3 - charges" each responsible for one type of hidden time signals.

5. "Gravity"

When placed near a huge bulk of matter, a huge congregation of emitting sources, a detector receives an enormous queue of scout signals. Though this queue exists in hidden time, it effectively slows down physical time, just like gravity does.

We are currently in the process of making quantitative estimations of this gravitational effect. These estimations will provide one possible test for the validity of our model.

## 6.     Conclusion

We suggest a very simple underlying picture – a deep structure – explaining *the origin* of quantum electro-dynamical amplitudes.

Though this model is based on very simple idea, its principles generate many subtle details that seem at first glance to overcomplicate the model. However, many of these details turn out to have direct analogies in mainstream physics.

Some aspects of our hidden–time conversational model are very close to those developed by Prof. Lewis E. Little [5], but Little's work does not use hidden time. On the other side, Prof. Alex Kaminskii from Tbilisi shares our hidden time idea [12, 13], but he does not introduce explicit talks with reverse propagation of signals.

Combining hidden time and reverse propagation is fundamental novelty of conversational model.

Though the conversational model and its proposal of hidden time are far from conventional, we believe these concepts open explanatory possibilities that are far too important to ignore.

Some results of the paper were reported at 11[th] International Conference on Composites and Nano Materials (ICCE – 11, 2004, South Carolina, USA) [11].

## *References*

1. *R. P. Feynman*. "QED – a strange theory of light and matter". *Princeton, New Jersey: Princeton University Press, 1985*.
2. *R. P. Feynman, A. Hibbs*. "Quantum mechanics and path integrals". *McGraw-Hill Book Company. New York 1965.*
3. *J. Cramer.* "Transactional Interpretation of Quantum Mechanics". *Reviews of Modern Physics 58, 647-688, June 1986*.
4. *H. Salacker, E. P. Wiegner*. "Quantum Limitations of the Measurement of Space – Time Distances". *Physical Review, Vol. 109, No. 2, Jan. 15, 1958*.
5. *Lewis E. Little.* Theory of Elementary Waves. *Physical Essays, vol. 9, No. 1, 1996*.
6. *A. Yariv.* Quantum electronics. *John Wiley and Sons, New York (1989)*.
7. *Karl von Frisch*. "The Dance Language and Orientation of Bees". *Cambridge: The Belknap Press of Harvard University Press, 1967.*
8. *Thomas D. Seeley.* "Honeybee Ecology: A Study of Adaptation in Social Life". *Princeton, NJ: Princeton University Press, 1985.*
9. *P. V. Kurakin.* "Hidden variables and hidden time in quantum theory". *Keldysh Institute of Applied Mathematics (RAS) preprint No. 33, 2004.*
10. *P. V. Kurakin, G. G. Malinetskii.* "The hidden time concept and quantum electrodynamics". *Electronic magazine "Quantum Magic" (http://quantmagic.narod.ru), 2004, vol. 1 No. 2, pp. 2101 – 2109*.
11. *P. V. Kurakin, G. G. Malinetskii.* "Proposed Fundamental Procedure of Detecting Physical Time at Atomic Level and Quantum Theory Paradoxes". *A report at ICCE – 11, August 2004*.
12. *A. V. Kaminskii.* "Hidden space – time in physics". *Electronic magazine "Quantum Magic" (http://quantmagic.narod.ru), 2005, vol. 2, No. 1, pp. 1101 – 1125*.
13. *A. V. Kaminskii.* "On the hidden nature of spin". *Electronic magazine "Quantum Magic" (http://quantmagic.narod.ru), 2005, vol. 2, No 2, pp. 2114 – 2131*.

## Appendix A

Roughly 95% of the bees in a hive are rigid conformists. They're what the American essayist Henry David Thoreau would call, "the mass of men" who lead "lives of quiet desperation". Except they're not men, they're women. And desperation, when it hits them, is part of a search-and-swallow strategy. When day breaks and it's time for work, most of the foragers go along with the herd and fly out to the fashionable flower patch of the moment to pick up nectar, pollen, and water.

Then there are the self-indulgent few, the non-conformists, the angry young women, who refuse to go with the flow. Instead they take off on long flights of their own, going nowhere in particular, simply following their whims. What a waste of time and energy. What utter selfishness. What a Bohemian shunning of conformity and fashion.

But the time-wasting ways of the solitary rebels pay off in the end. They pay off, of all things, for the conformists. Eventually the old flower patch runs out of pollen and nectar. That's when quiet desperation sets in.

Just like you and me, bees need attention to keep their spirits high. Foragers get that attention on the lip of the hive, a lip that serves as a loading dock. On that dock are unloader-bees, the bees that grab the cargo from the foragers and shuttle it into the hive. Those unloaders are in close communication with the inside workers who know the hive's interior needs. The unloaders know when the colony is filled with pollen but needs more nectar, or has its fill of water and now needs something else. On hot days the bees working inside of the hive slap water on the wax walls of the wax hexagonal cells inside to air-condition the place. If the temperature inside the hive shoots too high, the pupae, the young still in their eggs, will die. On a hot day when a forager arrives with her side-and-leg pouches filled with water, the unloaders rush over and crowd around her, eagerly unpacking her cargo. She's treated like a superstar.

But once the hive is cooled, a bee that shows up with water is wasting her time. The unloaders ignore her as if she didn't exist. Think what being ignored when you arrive with what you think is a gift would do to you and me. At the very least it would depress you. It does the same thing to a bee.

Things get even worse if the fashionable flower patch of the moment is running out of pollen and nectar. When a worker bee sets off to follow the crowd and comes back with her cargo-pouches nearly empty, the unloaders pass her by as if she were dirt. Attention means everything in a world of fashion. The whole reason we go with the trend is to get others to look at us admiringly. That's true of you and me. And it's equally true of bees.

Bees who are shunned at the loading dock stagger around as if they are stunned—or more as if they have lost their sense of purpose, their sense of meaning. That's when the self-indulgence of the hippie explorer bees shows its worth. Crowds of discouraged forager bees wander around inside the hive looking for some way to lift their spirits, some way to entertain themselves. And they find it. Four or five of the explorer bees have accidentally bumbled into new flower patches or new water puddles.

And they are not shy about advertising their discoveries. Like street buskers or soapbox preachers, they dance their news. Four or five of them compete on the inside wall of the hive like break dancers vying for your attention at New York's Times Square. You've heard about these dances—the famous figure eights that spell out direction, distance, and wind speed on the way to the flower patch and back. It's a very complex language for a bee with only a tiny number of brain cells.

But the discouraged foragers gather round to watch the dancers flash and flaunt. Some explorers dance more enthusiastically than others. The most outrageous enthusiasts attract the biggest crowds. If the dance is sufficiently persuasive—which means if the bee dancing the message dances longer than her competitors, if she just won't give up—a few of the conformists will catch a bit of the dancer's enthusiasm and go out and check her report. If they're impressed with what they find, they come back and join the dance. If they're not impressed, they don't.

Eventually one bee manages to gather the biggest audience and the largest number of background dancers. That's when the hive makes up its mind. The conformists go off in a pack to mine the flower patch the winner of the dance contest advertised. When the foragers come back home with pouches full of stuff the hive needs, the unloaders rush to them, make a fuss over them, and unload them as quickly as they can. The foraging conformists get what they need most, attention. They sharpen up as if they have a sense of purpose, a sense of mission, and a sense of meaning again.

Meanwhile the explorer bees—the bee world's beatniks and hippies—go off on their self-indulgent flights and buzz off the beaten path, selfishly pursuing their trend-bucking rambles again.

The hive survives thanks to waste of time and energy. It survives thanks to the explorer's useless consumption of fuel. It survives thanks to self-indulgence. It survives thanks to a hippie luxury.

Then there's the waste of material goods it takes to make the bee system work. Remember, this isn't just a system of bees; it's a system of flowers, too. It's a system of exchange and commerce. Bees pollinate flowers, and flowers pay the bees with pollen and nectar. But, like the dancing explorer bees who promoted, marketed, and advertised their finds, flowers depend on advertising too. That advertising comes in the form of flamboyant material excess. It comes in the form of flower petals and intricately shaped passageways at a flower's heart. Flowers are billboards erected to attract the attention of bees." Howard Bloom. "In Praise of Consumerism: The Spiritual Fruits of Materialism". Lecture delivered at The Next Stage, New York, New York, February 5, 2005.

### Reverences to Appendix A

1. Thomas D. Seeley and Royce A. Levien. "A Colony of Mind: The Beehive As Thinking Machine." *The Sciences*, July/August, 1987: 38-42.
2. Thomas D. Seeley. *The Wisdom of the Hive: The Social Physiology of Honey Bee Colonies*. Cambridge, Massachusetts: Harvard University Press, 1995.
3. Howard Bloom. *The Lucifer Principle: a scientific expedition into the forces of history*. New York: Atlantic Monthly Press, 1995.
4. Howard Bloom and Michael J. Waller. "The Group Mind: Groups as Complex Adaptive Systems." Human Behavior and Evolution Society Annual Meeting, 1996.
5. Howard Bloom, (2000). *Global Brain: The Evolution of Mass Mind From The Big Bang to the 21$^{st}$ Century*. New York: John Wiley & Sons.

## Appendix B

The dynamics of ribs and nodes in our conversational hidden-time model are also very similar to the random search patterns of a centrosome in a cell's skeleton – the cytoskeleton. A centriole is spherical. It sends out tubules in random directions. Those tubules are temporary probes; they are the cytoskeleton's equivalent of scout bees. They are ribs made of tubulin, a protein that can polymerize or depolymerize easily, either melting away or gaining semi-permanence.

The new tubules that find a useful docking site, a useful destination, a kinetochore, are blessed with permanence. The docking site gives the hopeful rib's tubulin an extra protein that strengthens it against dissolution.

The unlucky microtubules—the unlucky ribs--that fail to find a docking point, a node that "needs" them, have no such luck. They dissolve rapidly. Why? Because they aren't wanted by the mesh they are a part of. They don't receive the protein signals that tell them they are necessary to the receiving social web.

Apoptosis, programmed cell death, follows a similar pattern. Cells stay alive while they sense that they are needed. When they no longer receive social signals telling them that their receivers need their services, they commit chemical suicide.

Microtubules in a cell and in the cells of an organism depend for their life and death on conversation with others, on the signals that tell them that the larger mesh of which they are a part needs them or finds them unnecessary. The deep structure underlying our conversational model for quantum particles also appears in the underlying mechanisms of life.

**References to Appendix B**

1.  Marc Kirschner and John Gerhart  July, 'Evolvability' in the Proceedings of the National Academy of Sciences, USA, (1998) 95, 8420-8427
2.  C.M. Payne, C. Bernstein, H. Bernstein. "Apoptosis overview emphasizing the role of oxidative stress, DNA damage and signal-transduction pathways." *Leukemia and Lymphoma*, September 1995: 43-93.
3.  David Tomei, Frederick O. Cope, editors. *Apoptosis: the molecular basis of cell death*. Plainview, NY: Cold Spring Harbor Laboratory Press, 1991.